# Low Temperature Phase Transitions of the Ionic Liquid 1-Ethyl-3-methylimidazolium Dicyanamide

Kalil Bernardino, Thamires A. Lima, Mauro C. C. Ribeiro*

*Laboratório de Espectroscopia Molecular, Departamento de Química Fundamental, Instituto de Química, Universidade de São Paulo, 05508-000 São Paulo, SP, Brazil*
* email: mccribei@iq.usp.br

**ABSTRACT**

Several calorimetric measurements have shown that 1-ethyl-3-methylimidazolium dicyanamide, $[C_2C_1im][N(CN)_2]$, is a glass-forming liquid, even though it is a low-viscous liquid at room temperature. Here we found slow crystallization during cooling of $[C_2C_1im][N(CN)_2]$ along Raman spectroscopy measurements. The low-frequency range of the Raman spectrum shows that the same crystalline phase is obtained at 210 K either by cooling or by reheating the glass (cold-crystallization). Another crystalline phase is formed at *ca.* 260 K just prior the melting at 270 K. X-ray diffraction and calorimetric measurements confirm that there are two crystalline phases of $[C_2C_1im][N(CN)_2]$. The Raman spectra indicate that polymorphism is related to $[C_2C_1im]^+$ with the ethyl chain on the plane of the imidazolium ring (the low-temperature crystal) or non-planar (the high-temperature crystal). The structural reason for the glass-forming ability of $[C_2C_1im][N(CN)_2]$, despite of the relatively simple molecular structures of the ions, was pursued by quantum chemistry calculations and molecular dynamics (MD) simulations. Density functional theory (DFT) calculations were performed for ionic pairs in order to draw free energy surfaces of the anion around the cation. The MD simulations using a polarizable model provided maps of occurrence of anions around cations. Both the quantum and classical calculations suggest that the delocalization of preferred positions of the anion around the cation, which adopts different conformations of the ethyl chain, is on the origin of the crystallization being hampered during cooling and the resulting glass-forming ability of $[C_2C_1im][N(CN)_2]$.

**I. INTRODUCTION**

Understanding the physico-chemical properties of ionic liquids in terms of the molecular structure of the constituent ions is needed for the growing scope of technological applications of these interesting systems.[1-5] Studies concerning phase transitions of ionic liquids, that are obviously important for practical purposes, have unveiled how ionic interactions and molecular conformation determine the local arrangement of ions in the liquid phase and the crystalline structure.[6-9] Calorimetric measurements have shown that ionic liquids may exhibit complicated patterns of phase transitions including crystallization, glass transition, solid-solid transitions, and complex pre-melting behavior.[10,11,12,13,14,15] At least three characteristic patterns of thermal behavior have been found by differential scanning calorimetry (DSC) of ionic liquids: *i.* crystallization; *ii.* supercooling and glass transition; *iii.* devitrification by reheating the glass (the so-called cold-crystallization). It should be noted that this is not a strict distinction, since a given ionic liquid undergoes crystallization or glass transition depending on slight change in cooling rate or different thermal history experienced by the sample.[16] Nevertheless, crystallization under cooling is expected when the molecular structures of cation and anion are relatively small and symmetric. For instance, the cation 1-ethyl-3-methylimidazolium, $[C_2C_1im]^+$ (see the molecular structure in Figure 1), combined with small anions such as $Cl^-$ or $Br^-$ gives solid salts at room temperature, and combined with anions such as $[NO_3]^-$, $[CF_3SO_3]^-$, or $[NTf_2]^-$ (bis(trifluoromethanesulfonyl)imide), gives low-viscous liquids that crystallize under cooling.[17,18,19] Longer alkyl chain in 1-alkyl-3-methylimidazolium cation results in a more viscous liquid and enhances the glass forming ability. Ionic liquids in which the cation has an amphiphilic molecular structure exhibit nanometric structural heterogeneities due to non-polar regions made of alkyl chains and polar regions made of anions and the polar part of cations.[20,21,22] As the alkyl side chain of the cation becomes very long, liquid crystals are obtained.[8,23]

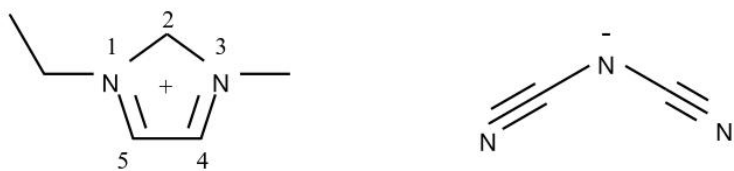

**Figure 1.** Molecular structures of 1-ethyl-3-methylimidazolium, $[C_2C_1im]^+$, and dicyanamide, $[N(CN)_2]^-$. Numbering of atoms of the imidazolium ring is given for further reference.

In the case of a very viscous liquid based on the relatively small $[C_2C_1im]^+$ cation, *e.g.* $[C_2C_1im][HSO_4]$, the glass forming ability follows as a consequence of the high viscosity.[14,24] Then the interesting issue is the reason of high viscosity, in the case of $[C_2C_1im][HSO_4]$ being the occurrence of anion–anion hydrogen bonded structures.[14,24] When the ethyl chain is replaced by a benzyl group (1-benzyl-3-methylimidazolium, $[BzC_1im]^+$) low viscosity liquids can be obtained, *e.g.* $[BzC_1im][N(CN)_2]$, which however does not crystallize under cooling.[25] The structural reason for the glass-forming ability of $[BzC_1im][N(CN)_2]$, despite of its low viscosity, was assigned to the steric hindrance due to the conformational flexibility of the bulky and anisotropic benzyl group that make ionic packing difficult.[25]

The title compound, 1-ethyl-3-methylimidazolium dicyanamide, $[C_2C_1im][N(CN)_2]$, has low viscosity at room temperature. (Viscosity values of $[C_2C_1im][N(CN)_2]$ at 298.15 K reported by different authors cover the range $14.5 < \eta < 17.4$ mPa.s).[26,27,28,29,30] Furthermore, the molecular structures of $[C_2C_1im]^+$ and $[N(CN)_2]^-$ are relatively simple in comparison to other more complex ionic liquids. These two features taken together give us the expectation that $[C_2C_1im][N(CN)_2]$ should crystallize under cooling. However, DSC data give no indication of crystallization of $[C_2C_1im][N(CN)_2]$ upon cooling, instead it undergoes glass transition.[31-35] Crystallization of $[C_2C_1im][N(CN)_2]$ was found only upon heating from the glassy phase back to the supercooled liquid range (cold-crystallization). For easy reference, Table 1 summarizes the temperatures of phase transitions of $[C_2C_1im][N(CN)_2]$ available in the literature. It is also manifest in Table 1 that there is no consensus in the literature whether there are two, one or none solid-solid transition.

**Table 1.** Temperature (in Kelvin) of glass transition ($T_g$), cold-crystallization ($T_{cc}$), solid-solid transition ($T_{ss}$), and melting ($T_m$) of [C$_2$C$_1$im][N(CN)$_2$] from the literature.

|           | $T_g$ | $T_{cc}$ | $T_{ss}$ | $T_m$ |
|-----------|-------|----------|----------|-------|
| Ref. [33] | 177   | 214      | 229, 245 | 255   |
| Ref. [31] | 169   | 216      | 223      | 252   |
| Ref. [35] | 182   | 226      | –        | 263   |
| Ref. [32] | 183   | 219      | –        | 261   |
| Ref. [34] | –     | 259      | –        | 268   |

The question that arises is the reason for [C$_2$C$_1$im][N(CN)$_2$] being a glass former. In this work, we use Raman spectroscopy to follow the phase transitions of [C$_2$C$_1$im][N(CN)$_2$] along slow cooling and heating cycles. We found crystallization of [C$_2$C$_1$im][N(CN)$_2$] during cooling, but crystal growing is such a very slow process that usual cooling rates in DSC measurements may preclude crystallization of being observed. It is shown that there are two crystalline phases of [C$_2$C$_1$im][N(CN)$_2$] with different conformations of the [C$_2$C$_1$im]$^+$ cation.

On the other hand, the glass forming ability of [C$_2$C$_1$im][N(CN)$_2$], despite of the relative simple molecular structures of the ions and the low viscosity of the liquid at room temperature, is an interesting issue. The structure of the ionic liquid [C$_2$C$_1$im][N(CN)$_2$] has been already investigated by molecular dynamics (MD) simulations, both classical and *ab initio* MD.[36,37,38] The MD simulations have shown that preferred positions for the [N(CN)$_2$]$^-$ anions spread all around the [C$_2$C$_1$im]$^+$ cation. Herein, we carried out calculations following two approaches. First we performed quantum chemistry calculations of ionic pairs in which [N(CN)$_2$]$^-$ is placed at different positions on a sphere of fixed radius around [C$_2$C$_1$im]$^+$. The set of calculated energies is used to build up a surface of free energy for anions around the cation. This methodology of quantum chemistry calculation of free energy surface is a useful tool to validate classical models for ionic liquids as it will be shown in a forthcoming paper. Then we performed classical MD simulations using a previously proposed polarizable model for [C$_2$C$_1$im][N(CN)$_2$].[39] Both the DFT and MD simulations result in consistent surfaces of free energy for anions around the cation and provide a hint for the reason of [C$_2$C$_1$im][N(CN)$_2$] being a glass forming liquid.

## II. EXPERIMENTAL DETAILS

The ionic liquid [$C_2C_1$im][N(CN)$_2$] was purchased from Iolitec (>98% purity) and used without further purification. On the other hand, it is important to work with a dry sample because the presence of water has a significant effect on the ionic interactions and consequently in the viscosity of the liquid and phase transitions. The ionic liquid was dried by leaving the sample for several hours under high vacuum ($10^{-5}$ mbar). The effectiveness of the drying procedure is easily confirmed by the absence of band due to water in the high frequency range of the infrared (IR) spectra. Figure S1 of the Supporting Information shows IR spectra of a [$C_2C_1$im][N(CN)$_2$] sample exposed to moisture and after *ca.* 1 hour in a Vertex 80v Bruker spectrometer whose sample chamber is kept under vacuum (~1 mbar). It is clear from Figure S1 that the moderate vacuum of the IR spectrometer is already enough to remove water. Thus, the high vacuum condition of the calorimetric, Raman, and X-ray diffraction measurements performed in this work guarantees that the sample is free from water contamination.

We used a Quantum Design Physical Properties Measurement System (PPMS) for the calorimetric measurements. The heat flow from the sample is measured as function of temperature after a heating or cooling ramp is established. A Peltier element attached to a blank puck is used as heat flow sensor which is also the sample support and the whole setup is kept under ~$10^{-4}$ mbar provided by the PPMS. Measurements were done along cycles of cooling and heating in the 190–300 K range under different rates (0.2, 0.5, and 1.0 Kmin$^{-1}$).

The Raman spectrometer used was a Horiba-Jobin-Yvon T64000 triple monochromator equipped with CCD. Raman spectra were excited with the 659.5 nm line of a solid-state Cobolt laser. The spectra were obtained in the 180° scattering geometry with no polarization selection of the scattered radiation. Spectral resolution was kept at 2.0 cm$^{-1}$. We used a micro-cryostat Janis ST-500 working with liquid nitrogen coupled to a temperature controller model 325 of LakeShore allowing for ±0.1 K of temperature control. A small drop of the ionic liquid lies on the copper sample mount plateau of the micro-cryostat working under high vacuum condition, so that the sample is not contaminated with water from ambient moisture. Raman spectra were obtained using the Olympus BX41 microscope coupled to the Horiba-Jobin-Yvon spectrometer. The Raman measurements followed different cooling and heating protocols as discussed in the next section. It took *ca.* five minutes for recording the Raman spectrum at each spectral window of interest (see Figures 3 and 4 below).

Temperature dependent X-ray diffraction measurements were performed for samples loaded in a copper holder. The measurements were taken within the $150 - 300$ K range using a Rigaku Rint-2000 equipment with a CuK$_\alpha$, $\lambda = 1.5406$ Å radiation. Samples were submitted to a cycle of cooling and heating using a He closed-cycle cryostat (Ulvac Cryogenics Mini Stat) under high vacuum. Uncertainty in temperature is ±0.5 K. Diffractograms were recorded from $5° \leq 2\theta \leq 40°$, with 0.02° step size and 5 s per step, in the same temperatures (210 and 260 K) that the Raman measurements indicated the formation of different crystalline phases of [C$_2$C$_1$im][N(CN)$_2$].

## III. COMPUTATIONAL DETAILS

The structures of the [C$_2$C$_1$im][N(CN)$_2$] gas phase ionic pair for the density functional theory (DFT) study were produced by the software Themis developed by Dr. F. M. Colombari to systematically perform rigid-body translations, rotations and precessions for a pair of molecules.[40] A spherical shell of radius 5.55 Å centered on the N3 atom of the cation (the nitrogen bonded to the methyl group, see Figure 1) was divided into a 92 points grid used for the translational movements of the anion (see Figure S2 in Supporting Information). The anion N(CN)$_2$$^-$ was placed with its central nitrogen atom over one point $t$ of this grid and 42 rigid body rotations were performed using a 42 points spherical grid now centered over the anion. For each rotation $r$, 8 precession movements $p$ were also performed. The process is repeated for the anion in each point $t$ at the translation grid in order to produce an ensemble of ionic pair structures. Thus, a total of 30912 structures for the ionic pair were generated, but the ones with significant atom superposition were discarded, so that about half this number of ionic pairs were actually computed.

The energy $E(t,r,p)$ of each ionic pair was computed by single point CAM-B3LYP/Def2-TVZPD[41] calculations with D3BJ correction[42] for dispersion with the Orca 4.0.0.2 software.[43] The free energy of the translational point $t$, $F(t)$, is calculated by the standard expression in the canonical ensemble:



$$F(t) = -k_B T \ln \left( \sum_{r,p} exp \left[ -E(t,r,p)/k_B T \right] \right) - F(t'),$$

(1)

where the summations run over all of the rotation and precession points associated to each point $t$, $k_B$ is the Boltzmann constant, and $T$ is the temperature. The $F(t')$ is the smallest free energy value, *i.e.* $F(t')$ sets the zero of the scale to the most favorable point at the translation grid. Since there are two important conformations for the cation, one with the ethyl group at the same plane of imidazolium ring (planar $[C_2C_1im]^+$) and the other out of the plane (non-planar $[C_2C_1im]^+$), free energy surfaces were calculated for both of these cation conformations.

Classical MD simulations of liquid $[C_2C_1im][N(CN)_2]$ with 300 ions pairs were performed with the LAMMPS package.[44] The initial configuration was produced using Packmol[45] and fftool.[46] The Drude polarizable forcefield proposed by Padua,[39] built on a previous non-polarizable model,[47] was used to describe pair interactions. Two thermostats were employed:[48] the atoms were kept at the temperature of 300 K with a damping time of 200 fs, and the movement of the Drude particles in relation to the atomic cores were thermalized at 1 K with damping time of 50 fs in order to keep them close to the equilibrium position for each structure of the atomic nucleus. Nosé-Hoover thermostat were employed for both the Drude particles and the atoms core.[49] Nosé-Hoover barostat was applied for the atoms cores in the NPT ensemble simulations with $P = 1$ bar. A cutoff of 12 Å was used for both electrostatic and dispersion interactions, with the PPPM scheme to correct the effect of long-range electrostatic interactions. Periodic boundary conditions were applied in the three dimensions in order to reduce small systems artifact. An integration time step of 1.0 fs was used in all of the simulations.

A total of six MD simulations of $[C_2C_1im][N(CN)_2]$ were performed. The first was a 125 ns NPT simulation with no restraint over the ethyl group conformation except by the torsion potential already implemented in the forcefield. In order to verify the effects of the ethyl group conformation over the structure and the dynamics of the liquid phase, two others 100 ns NPT simulations were performed with harmonic potentials with force constants of 40.0 kcalmol$^{-1}$degree$^{-2}$ centered at 0 or at 110 degree in order to keep the $[C_2C_1im]^+$ conformation planar or non-planar, respectively. This procedure



leads to two artificial liquids in which there are only planar or only non-planar $[C_2C_1im]^+$ conformers, instead of exhibiting a mixture of cation conformations.

When the $[C_2C_1im]^+$ was restrained to planar or non-planar conformation, the system evolved to liquids of smaller densities than the liquid without the restraints. This implies the issue whether any effect on the liquid dynamics upon the application of dihedral restraint is a simple consequence of the smaller density or a structural effect on ionic mobility. To tackle this issue, two new simulations in NVT ensemble were performed with the restraint applied to the dihedral angle, but with the volume equal to the average value found in the simulation without the restraint. Finally, in order to evaluate if the effect of the rigid dihedral on dynamical properties would hold even if there is a mixture of conformations, a sixth simulation of 40 ns in NVT ensemble was done starting from the equilibrated configuration of the simulation without restraints, but now including the harmonic potentials to hold each cation in its starting conformation. This procedure results in another artificial liquid in which each cation is unable to change the ethyl group conformation, but the distribution of dihedral angles is representative of the realistic (without restraint) liquid. This last system will be refereed as "mixed rigid". In every simulation the first 10 ns were discarded from the analyses in order to ensure the liquid equilibration.

The Travis[50] software was used for structural analyzes and dimer lifetime calculations of the liquid phase and graphical representations of structures. Spatial distribution functions and free energy surfaces were rendered with VMD 1.9.3.[51]

## IV. RESULTS AND DISCUSSION

### A. Phase Transitions

Figure 2 shows the calorimetric results obtained in the PPMS along cooling and heating of $[C_2C_1im][N(CN)_2]$ at a rate of $\pm0.2$ Kmin$^{-1}$. The exothermic peak at 233 K in the cooling curve (black line) indicates that crystallization of $[C_2C_1im][N(CN)_2]$ may occur during cooling in contrast with previous reports.[31-35] However, along the Raman spectroscopy measurements, where visual and microscopic images of the sample are available, we found that continuous cooling and heating at typical rates lead to partial crystallization as discussed below. Thus, the peak at 233 K seen in the cooling curve of Figure 2 most probably is related to partial crystallization. The heating curve (red line)



shows a single solid-solid transition at $T_{ss} = 253$ K, which should be compared to data from literature given in Table I. The heating was stopped at 257 K, the sample was cooled back to 240 K, and then the heating scan was continued (the region in magenta in Figure 2). The passage by the second time in this small cycle around $T_{ss}$ did not exhibit the peak suggesting that it is an irreversible process. Melting of $[C_2C_1im][N(CN)_2]$ is seen in Figure 2 at $T_m = 266$ K.

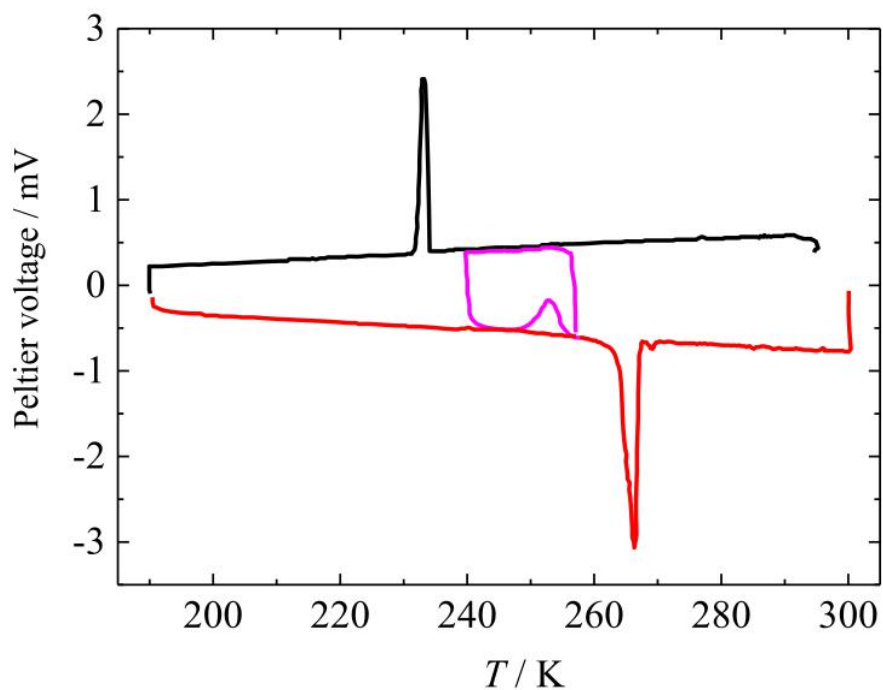

**Figure 2.** Calorimetric measurements of $[C_2C_1im][N(CN)_2]$ along cooling (black line) and heating (red line) at rate ±0.2 Kmin$^{-1}$. Exothermic peaks point upward. The cycle seen in the 240–260 K range (magenta line) corresponds to the heating scan up to 257 K, cooling back to 240 K, and then continuing the heating scan.

The thermal behavior of $[C_2C_1im][N(CN)_2]$ indicated by the calorimetric results shown in Table I and Figure 2 deserves a better account of the dependence of transitions with the thermal history experienced by the sample. We performed different cooling and heating protocols in the Raman spectroscopy study: step-wise temperature variation (steps of 5 or 10 K followed by *ca.* 30 min for equilibration at each temperature), continuous cooling (~1 Kmin$^{-1}$), and keeping the temperature fixed at specified values allowing for full crystal growth.



The glass is obtained by a continuous cooling of the sample down to temperature below the expected glass transition ($T_g \sim 175$ K, see Table 1). In line with previous reports, we also found cold-crystallization at 210 K by reheating the glassy phase. During cooling of $[C_2C_1im][N(CN)_2]$, we found it is indeed easily supercooled in the temperature range down to 210 K, as no crystallization was observed when the temperature is kept constant at a given value at least for *ca.* 30 min. When the temperature reached 210 K, in a few minutes white opaque crystal nuclei appeared in the sample. It should be noted that this temperature is close to the glass transition, so that the high viscosity of the liquid at such low temperature implies a very slow process of crystal growth. It took one hour for the crystalline phase to fulfill the small drop of the ionic liquid sample. After the complete formation of this crystalline phase, we increased the temperature stepwise looking for any solid-solid transition. The white opaque crystalline phase starts transformation to a limpid crystal in the range 250–260 K. It takes also *ca.* one hour to complete the solid-solid transition. It is worth noting that the solid-solid transition takes place close to the melting temperature (melting is a fast process at 270 K, but melting of the borders of the crystals is already evident at 265 K). Thus, the solid-solid transition might not be discernible during a continuous heating in a calorimetric measurement (see Table 1).

The left panel of Figure 3 shows the low-frequency Raman spectra of different phases of $[C_2C_1im][N(CN)_2]$. The crystal formed during cooling at 210 K (crystal I) and after the solid-solid transition at 260 K (crystal II) have different patterns of lattice vibrations. Interestingly, the high-temperature crystal II has a soft mode at 12 cm$^{-1}$ that is very intense in the Raman spectrum. The low-frequency Raman spectrum of the glassy phase of $[C_2C_1im][N(CN)_2]$ is also shown in Figure 3. The Raman spectrum of the glass exhibits the broad shape expected for the spectrum of an amorphous solid, including the so-called boson peak at *ca.* 16 cm$^{-1}$ that is a characteristic feature of Raman spectra of glasses.[52,53,54,55] It is clear from the photographs of the optical microscopy images shown in Figure 3 that the two crystals have different morphologies: the crystal II has needle shape, whereas crystal I exhibits an ill-defined shape.



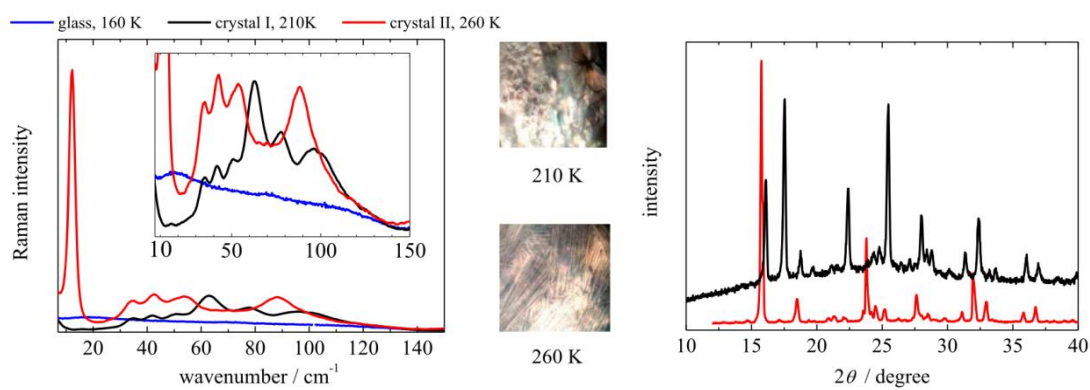

**Figure 3.** Left: Low-frequency Raman spectra of [C₂C₁im][N(CN)₂]: glass (160 K, blue line), crystal I (210 K, black line), and crystal II (260 K, red line). The inset shows the same data in other scale of Raman intensity. The intensities are normalized by the intensities of Raman bands in the high frequency range. Middle: Images of the optical microscopic of the Raman spectrometer for the crystal I (210 K, top) and crystal II (260 K, bottom). Right: X-ray diffraction patterns for the crystal I (210 K, black line), and crystal II (260 K, red line).

The right panel of Figure 3 shows X-ray diffractograms of crystals I and II. The X-ray diffraction measurements were performed following the same protocol used in the Raman measurements. The sample was cooled stepwise to 210 K, the temperature kept constant for about one hour while fast X-ray scans were done to monitor the full formation of crystal I. After recording the diffractogram of crystal I, the sample was heated stepwise and again waiting for full formation of crystal II at 260 K. It is clear from Figure 3 the different X-ray diffractions patterns of crystals I and II. The broad base line in the diffractogram of the low-temperature crystal I indicates that part of the sample remained amorphous.

We found reproducibility of the Raman spectra of crystals I and II obtained in different measurements as long as time is allowed for growing of each crystal as mentioned above. It is worth stressing that the Raman spectrum of crystal formed during cooling is exactly the same as the spectrum of the crystal I formed by cold-crystallization at 210 K (spectrum not shown). There is a large literature concerning Raman spectra of crystalline phases of ionic liquids obtained by cold-crystallization after heating the glass.[6,7,9] However, in the authors knowledge, [C₂C₁im][N(CN)₂] is the



first ionic liquid that is confirmed that the same crystalline phase is formed either by cold-crystallization upon heating of the glass or by cooling from the normal liquid phase.

Vibrational spectroscopy (Raman and infrared) has been a powerful tool to unveil the molecular conformations that the ions adopt in a given crystalline phase of ionic liquids.[6,7,9] In the case of $[C_2C_1im]^+$, two limiting conformations have the ethyl group in the plane or out of the plane of the imidazolium ring. The molecular structures of the conformers, hereafter called planar and non-planar, are shown in Figure 4. Quantum chemistry calculations of the vibrational frequencies and Raman and infrared intensities of $[C_2C_1im]^+$ indicated some spectral regions that could be used to characterize each of the conformers.[56,57] Figure 4 shows Raman spectra within the 340–500 $cm^{-1}$ range, which is very appropriated to distinguish the planar and non-planar $[C_2C_1im]^+$ conformers. The DFT calculations of vibrational frequencies were performed in Refs. [56] and [57] for the isolated $[C_2C_1im]^+$ cation. Herein, we confirmed these spectral patterns for the two conformers of $[C_2C_1im]^+$ along the quantum chemistry calculations of free energy surfaces of ion pairs to be discussed in the next section. For completeness, Figure S3 of Supporting Information shows the atomic displacements obtained by DFT calculations for the two vibrational modes related to the intense Raman bands in the 340–500 $cm^{-1}$ range. The eigenvectors shown in Figure S3 confirm the assignment of these bands to CH2-(N) bending vibrations as proposed in Ref. [56]. Comparison of experimental spectra with the prediction of quantum chemistry calculations indicates that crystals I and II contain $[C_2C_1im]^+$ in planar and non-planar conformations, respectively. Umebayashi *et al.*[56] provided a review in the Cambridge Structural Database of 35 crystalline structures of salts based on the cation $[C_2C_1im]^+$, most of them with the non-planar $[C_2C_1im]^+$ conformer (see figure 1 in Ref. [56]). The interesting result in Figure 4 is that $[C_2C_1im][N(CN)_2]$ has crystalline phases each one with a different $[C_2C_1im]^+$ conformer.



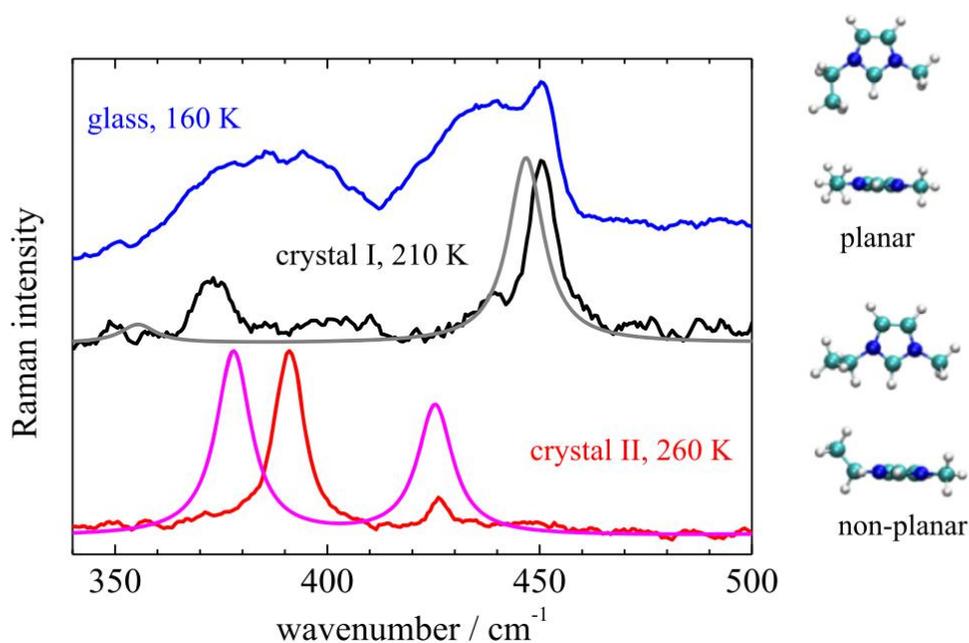

**Figure 4.** Experimental Raman spectra of [C$_2$C$_1$im][N(CN)$_2$] in the glassy phase at 160 K (blue line) and crystalline phases at 210 K (crystal I; black line) and 260 K (crystal II; red line). On the corresponding experimental data at 210 and 260 K, Raman spectra are shown as predicted for [C$_2$C$_1$im]$^+$ in planar (gray line) and non-planar (magenta line) conformation according to the frequencies and intensities calculated by Lassègues *et al.*[57] using DFT B3LYP/6-311G*. Theoretical spectra were modelled by Lorentzian bandshapes with 5 cm$^{-1}$ of bandwidth. The spectra were vertically shifted for help visualization. Molecular structures are shown for [C$_2$C$_1$im]$^+$ with the ethyl group in planar or non-planar conformation in relation to the plane of the imidazolium ring. Two views are shown for each [C$_2$C$_1$im]$^+$ conformer.

The Raman spectrum of the glassy phase shown in Figure 4 exhibits broad bands in the 340–500 cm$^{-1}$ indicating that there is a distribution of different [C$_2$C$_1$im]$^+$ conformers in the amorphous solid. Umebayashi *et al.*[56] used the temperature dependence of the relative intensities of Raman bands assigned to planar and non-planar conformers to estimate the enthalpy of the conformational change, $\Delta H$. They obtained $\Delta H \approx 2$ kJ/mol for the equilibrium non-planar $\leftrightarrows$ planar in [C$_2$C$_1$im][BF$_4$] and [C$_2$C$_1$im][CF$_3$SO$_3$], *i.e.* the non-planar is slightly more stable that the planar conformer. Lassègues *et al.*[57] estimated from the Raman spectra of [C$_2$C$_1$im][NTf$_2$] that 87% of the cations are in the non-planar conformation in the liquid phase at room temperature. The



Raman spectrum of $[C_2C_1im][N(CN)_2]$ in the spectral range shown in Figure 4 for the glassy phase is essentially the same as the spectrum of the liquid phase at room temperature (spectrum not shown). Therefore, the population distribution of $[C_2C_1im]^+$ conformers in the liquid phase is kept frozen in the glassy phase.

Summing up, the ionic liquid $[C_2C_1im][N(CN)_2]$ is indeed a good glass-forming liquid, but crystallization can be observed during cooling. If the low-temperature crystal I is heated before the slow process of crystal growth being complete, we found that a mixture of crystals I and II is formed. This is clearly indicated by the low-frequency Raman spectra shown in Figure S4 of the Supporting Information. The Raman spectra recorded during stepwise heating from 210 to 260 K is a superposition of the spectra of crystals I and II, the latter contribution increasing with temperature. In previous publications, Raman[58] and infrared[59] spectra of $[C_2C_1im][N(CN)_2]$ have been discussed in the spectral range of the symmetric and antisymmetric $C\equiv N$ stretching modes of the anion, $\nu_s(CN)$ and $\nu_{as}(CN)$. It is expected that these vibrations act as probes of the local environment experienced by the anion. For completeness, Figure S5 of the Supporting Information shows the 2110–2220 cm$^{-1}$ range of Raman spectra for the liquid, crystal I, and crystal II phases of $[C_2C_1im][N(CN)_2]$. The vibrational frequencies of $\nu_s(CN)$ and $\nu_{as}(CN)$ shift by few wavenumbers in the different thermodynamic states of $[C_2C_1im][N(CN)_2]$. The frequency shift among crystal I and II is particularly clear for the $\nu_{as}(CN)$ mode, whereas the broad $\nu_s(CN)$ and $\nu_{as}(CN)$ bands in the liquid phase indicate that there is a multiplicity of arrangements of $[N(CN)_2]^-$ around the cation in the liquid phase. However, the experimental frequency shifts are small, so that quantum chemistry calculations do not give an unequivocal correlation between ion pair structure and calculated $\nu_s(CN)$ and $\nu_{as}(CN)$ frequencies. Thus, Raman spectroscopy indicates that there are different conformations of $[C_2C_1im]^+$ and arrangements of $[N(CN)_2]^-$ anions around the cation in the liquid phase, but a more detailed picture of the interplay between conformation and local structure demands the quantum chemistry calculations and MD simulations to be discussed in the next section.



**B. Liquid Structure**

The local structure of $[C_2C_1im][N(CN)_2]$ has been investigated by DFT calculations of ionic pairs[60] and by classical and *ab initio* MD simulations of the liquid phase.[36,37,38] In the quantum chemistry study of ionic pairs, Izgorodina *et al.*[60] put forward the interesting proposition that the melting temperatures of pyrrolidinium and imidazolium based ionic liquids correlate to the ratio of the total binding energy to the dispersion energy contribution. The rational for this proposition is that the enthalpy of fusion should be proportional to the total interaction energy, whereas the entropy of fusion should be proportional to the dispersion component. Furthermore, they proposed that transport coefficients (conductivity and viscosity) correlate to the dispersion interaction energy.[60] A comparison of the total and dispersion energies calculated for ionic pairs of 1-alkyl-3-methylimidazolium cations with different anions indicated that $[N(CN)_2]^-$ results in relatively large contribution of dispersion (-51.9 kJmol$^{-1}$) to the total binding energy (-361.3 kJmol$^{-1}$).[60] In this case, the dispersion contribution is relatively large because both cation and anion are electron-density rich species. A particularly important finding from these DFT calculations for the present discussion is the fact that $[N(CN)_2]^-$ has different sites for interaction and also different ion pair structures of local minimum energy are possible in which the anion lies above/below the plane or in the plane of the imidazolium ring of $[C_2C_1Im]^+$.

The physical picture of ion pair structures obtained from the DFT calculations is manifest in the local structure obtained from MD simulations of the liquid. Classical MD simulations using a non-polarizable model[36] provided a detailed account of the static structure factor obtained by X-ray diffraction measurements for a series of $[C_2C_1Im]^+$ based ionic liquids with different cyano-anions ($[SeCN]^-$, $[SCN]^-$, $[C(CN)_3]^-$, $[B(CN)_4]^-$, and $[N(CN)_2]^-$ ). The distribution of small anions around the $[C_2C_1Im]^+$ cation is more localized about the hydrogen atoms in the plane of the imidazolium ring, and the distribution above/below the imidazolium rings increases with the anion volume. In the case of $[N(CN)_2]^-$, the distribution of anions spreads on the top and on the plane, mainly along the direction of the more acidic hydrogen atom H2 of the imidazolium ring.[36] In the *ab initio* MD simulations of $[C_2C_1Im][SCN]$, $[C_2C_1Im][N(CN)_2]$, $[N(CN)_2][C(CN)_3]$, and $[N(CN)_2][B(CN)_4]$, Weber and Kirchner[38] emphasizes another consequence of the anion size on the liquid structure. The small anion $[SCN]^-$ interacting with the cation in the plane of the imidazolium ring allows for cation–cation



$\pi$–$\pi$ interaction, whereas bulkier anions occupying the top of the ring compete with the cation–cation interaction. Even though the preferred position of $[N(CN)_2]^-$ is on-top of the imidazolium ring, $[N(CN)_2]^-$ is also involved in directional hydrogen bonds between its terminal nitrogen atom and the hydrogen atoms of the imidazolium ring.[38]

The DFT calculations of ionic pairs and classical MD simulations with a polarizable model performed in this work emphasize the interplay between the conformation of $[C_2C_1im]^+$ (planar and non-planar conformers, see Figure 4) and the distribution of $[N(CN)_2]^-$ around the cation. It is worth noting that the DFT calculations were not performed for few optimized ionic pair structures as in Ref. [60] (see the four structures of $[C_2C_1im]^+/[N(CN)_2]^-$ pairs in figure S3 of Ref. [60]). Herein, DFT calculations for many configurations of ion pairs generated a sphere of free energy for $[N(CN)_2]^-$ around $[C_2C_1im]^+$ as explained in Section III.

Figure 5 shows the free energy surfaces computed for the anion translation around the cation in the planar (a and b) and non-planar conformation (c and d). The radius of the sphere corresponds to a distance of 5.55 Å between the N3 atom of the cation and the central nitrogen atom of the anion. This distance was chosen on the basis of the corresponding radial distribution function calculated by MD simulation of the liquid phase. (Figure S6 of the Supporting Information shows this radial distribution function. The shoulder at 5.55 Å includes the region close to the hydrogen atom H2 bonded to the C2 atom. We focused on this spatial range because it is the region in which the density of the anion responds to the conformational change as discussed below with the MD results of Figure 6). The red regions correspond to low free energy values, *i.e.* favorable regions for the anion, whereas the blue regions correspond to high free energy, *i.e.* unfavorable regions. The DFT calculations for isolated ionic pairs shown in Figure 5 demonstrate that the preferred positions of anions around the cation respond to the conformation of the ethyl group. For the planar conformation of $[C_2C_1im]^+$, the most favorable region found for the anion is directly in front of the H2 atom (Figures 5.a and 5.b). In the non-planar conformation, the most favorable position slightly moves to the side of this hydrogen (Figure 5.c and 5.d). This effect is due to steric hindrance: when the ethyl group is out of the plane, the anion can approach the hydrogen by the same side of the ethyl group.



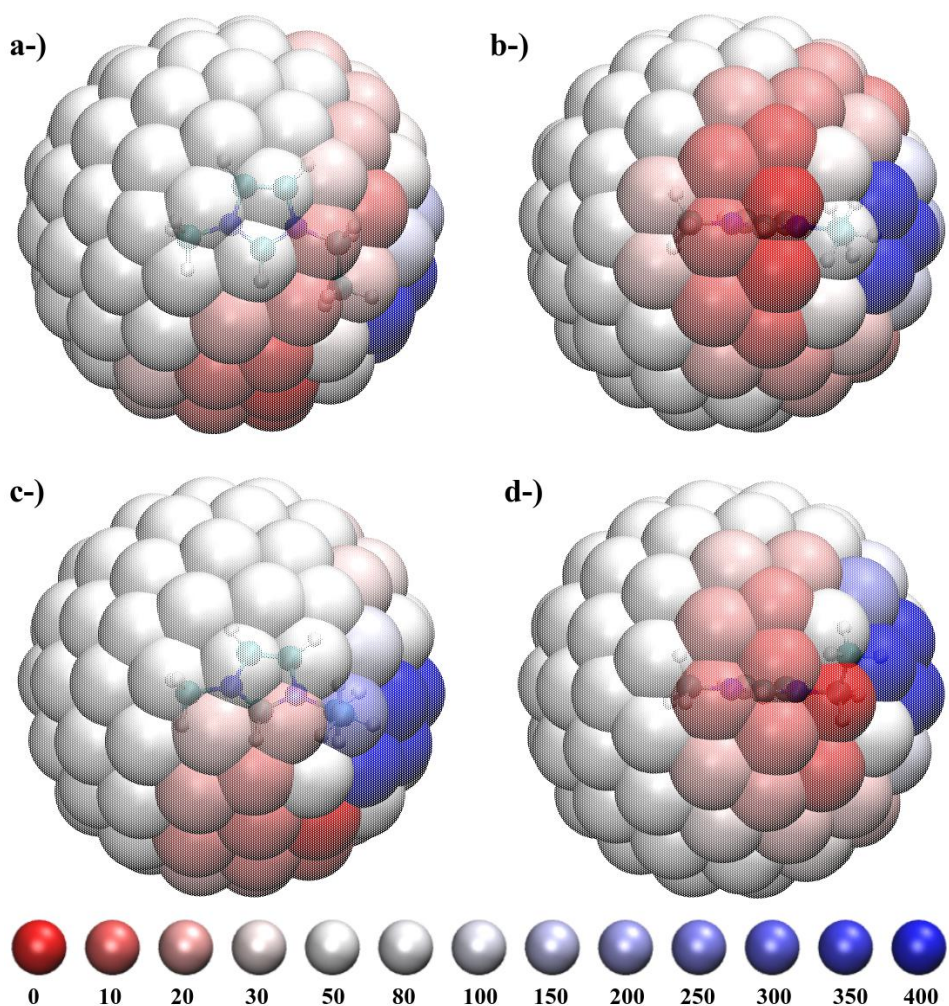

**Figure 5.** Free energy surfaces for the [C$_2$C$_1$Im][N(CN)$_2$] ionic pair computed by CAM-B3LYP/Def2-TVZPD calculations in a spherical grid of radius 5.55 Å around the N3 atom of the cation. Two different views are shown of the surface for each [C$_2$C$_1$im]$^+$ conformer: a-) and b-), planar conformer; c-) and d-) non-planar conformer. Color scale of free energy values (kJmol$^{-1}$) is given in the bottom.

In the molecular dynamic simulations of the liquid phase, the expected mixture of cation conformations was found (see the dihedral distribution in Figure S7 of Supporting Information), the non-planar being the dominant conformer. In order to understand the role of ethyl group conformation on the liquid structure, a new set of simulations were done with additional restraints to keep the [C$_2$C$_1$im]$^+$ fixed in planar or non-planar conformation as described in Section III. It is a common practice in MD simulations of ionic liquids to report the local structure by spatial distribution function



(SDF) of the distribution of ions around a central one.[36,37,38] The SDFs shown in Figure 6 indicate that the distribution of anions around cations change significantly depending on the cation conformation. In both the cases, the anion presents a large density above/below the imidazolium ring mainly close to the carbon 2 (Figure 6.a, c and e; for atom numbering, see Figure 1), but the distribution of the anion at the plane of imidazolium showed remarkable differences. For the planar conformation, there is a larger density of anions immediately in front of C2-H2 bond (Figure 6.b), but when the cation is forced to the non-planar conformation the density migrates to the sides of H2 (Figure 6.d) with some asymmetry, being larger close to the $CH_3$ side instead of the $C_2H_5$ side. The SDFs from the NPT simulations with restrained conformations are similar to the ones presented here for the NVT simulations, whereas the SDF for the system with a distribution of conformations of rigid cations (the "mixed rigid" system, results not shown) is very similar to the one found for the system without restraint.

In the realistic case with the ethyl group able to change conformation, the anions density behaves as intermediate between the simulations with restrained conformations (Figure 6.e and f), with medium densities both in front and at the sides of the C2-H2 bond, but with a larger accumulation at the side closest to the $CH_3$ group. A pertinent question is whether the difference observed for the two conformations of the cation is a many-body effect, in the sense that it happens only if every cation assumes the same conformation, as done in the simulations with restraints, or a local effect as the anions density responds instantly to the conformation of the closest cation despite the conformations of the remaining ones. This issue was addressed by the calculation of another set of SDFs from the simulation without restraints, but now picking as reference cations only those with a specific conformation at each time-step. The results obtained by this procedure are essentially the same obtained in the simulations with all cations in the same conformation. In other words, if a cation is found at the planar conformation in the liquid phase, the anion will present a large probability to be found in front of C2-H2 bond instead of on the sides of this bond. In contrast, if the cation assumes a non-planar conformation, there will be a smaller probability of finding the anion in front of C2-H2 bond and a higher probability to find it on the sides of that bond, regardless of the conformation of the remaining cations in the system. Since it is a local effect, it is reasonable that the distribution of the anion with a mixture of conformations (Figures 6.e and f) presents itself as an average over the distributions founded in the extreme cases (Figures 6.a to d).



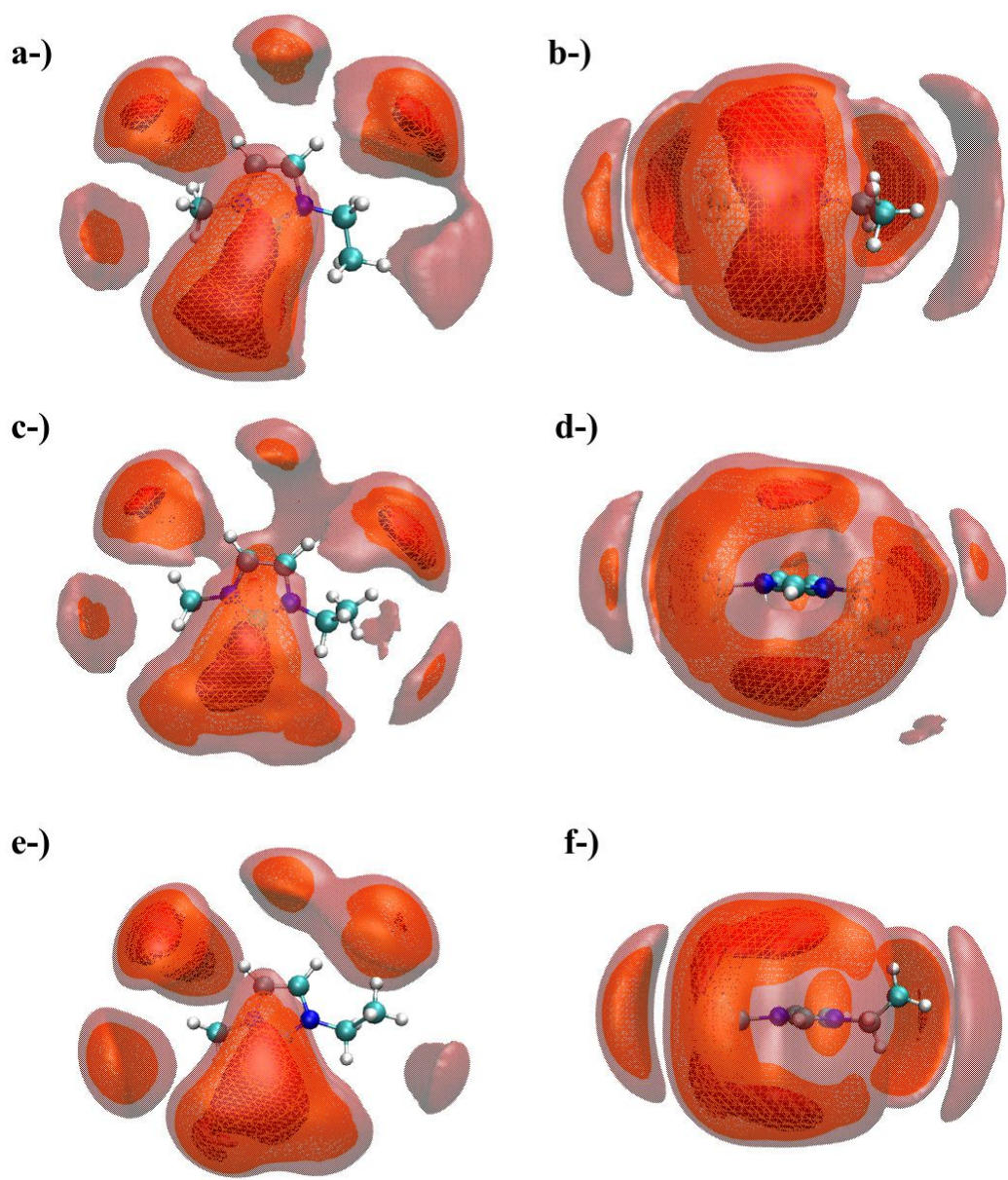

**Figure 6.** Spatial distribution function (SDF) of anions around the cation. Two different views are shown for each simulation, the ones at left displaying a view above the plane of imidazolium and at right with hydrogen atom H2 pointing to the reader. Panels a-) and b-): NVT simulation with planar $[C_2C_1im]^+$ conformer; c-) and d-): NVT simulation with planar $[C_2C_1im]^+$ conformer; e-) and f-) NPT simulation without restraints on the $[C_2C_1im]^+$ conformation. Isovalues: 8 nm$^{-3}$ (translucid red), 10 nm$^{-3}$ (orange wireframe) and 13 nm$^{-3}$ (solid red surface).



For completeness, we discuss how the liquid dynamics is sensitive to the conformational restraints. Table 2 displays the diffusion coefficients for both ions computed by the mean square displacement using Einstein relation, the viscosity computed by Green-Kubo method,[61] and the ionic pair lifetime. As mentioned in Section III, when performing an NPT simulation forcing all molecules to the same conformation, the density of the liquid decreases in relation to the realistic model without restraints. The reduction on density increases ionic mobility even in the absence of any structural change. However, the density effect by itself cannot explain totally the changes in the transport properties. The NVT simulations with density maintained the same as the unconstrained case also give significant increase of diffusion coefficients and decrease in ion pair lifetime and viscosity. This is a counterintuitive result: when the flexible group of the cation was forced to be rigid, the liquid becomes more fluid. A possible cause is that the conformational changes of the ethyl group is much faster than ionic diffusion. As the anion distribution responds to these frequent conformational changes of the cation, the movement of the ethyl group induces a correlated local movement of the anions, that is a way of dissipating energy in a flow, contributing to a higher viscosity in comparison to the artificial case where the movement of the ethyl group is restricted.

It should be noted that the faster dynamics upon restraint of ethyl group holds even if the system presents a mixture of conformations ("mixed rigid" system in Table II). Since the effect of cation conformation over the anion structure is essentially local, the similarity between results for the "mixed rigid" and the NVT system with every cation in the non-planar conformation is expected from the fact that the non-planar is the predominant conformer in the unrestrained system (see Figure S7). The non-planar conformer is the predominant in the "mixed rigid" system as the distribution of cations in the later was taken from an equilibrated snapshot from the unrestrained system.



**Table 2.** Density, diffusion coefficient of cation ($D_+$) and anion ($D_-$), dimer lifetime, and viscosity ($\eta$) calculated by MD simulations of [$C_2C_1Im$][$N(CN)_2$] at 300 K. Data shown correspond to MD simulations without restraint on the cation conformation (NPT ensemble), with every cation restricted in a conformation representative of the non-restrained case ("mixed rigid" system), and with [$C_2C_1Im$]$^+$ restrained in planar or non-planar conformation both in NVT and NPT ensembles.

| | density (g/cm$^3$) | $D_+$ ($10^{-10}$ m$^2$s$^{-1}$) | $D_-$ ($10^{-10}$ m$^2$s$^{-1}$) | dimer lifetime (ps) | $\eta$ (mPa.s) |
|---|---|---|---|---|---|
| no restrain, NPT | 1.097 | 1.5 | 1.8 | 1257 | 12.3 |
| mixed rigid, NVT | 1.097 | 11.2 | 12.9 | 191 | 5.4 |
| planar, NVT | 1.097 | 7.0 | 9.0 | 255 | 7.0 |
| non-planar, NVT | 1.097 | 10.5 | 12.1 | 195 | 4.7 |
| planar, NPT | 0.976 | 20.3 | 19.7 | 116 | – |
| non-planar, NPT | 0.923 | 28.3 | 32.7 | 74 | – |

The conclusions drawn by the SDFs shown in Figure 5 can be put forward as free energy surfaces allowing for a more direct comparison with the free energy surfaces obtained from the DFT calculations of ionic pairs. The density distribution of anions around cations can be used to generate a free energy surface analogous to the DFT results shown in Figure 5. The free energy at a given point $i$ around the cation, $F_i$, is calculated from the density of anions, $\rho_i$:

$$F_i = -k_B T \ln \left( \frac{\rho_i}{\rho_{max}} \right) . \quad (2)$$

The $\rho_{max}$ stands for the largest density and defines the arbitrary zero of the free energy as being the most favorable point for finding the anion. In analogy to the DFT calculations of ionic pairs, we set a color scale over a spherical surface around the cation for a distance of 5.55 Å around the N3 atom of the cation.

The red region in each surface of Figure 7 corresponds to the lowest free energy values computed over that spherical surface, whereas the blue region corresponds to the highest free energy values. As in the case of SDF (Figure 6), no significant difference is observed in the region above/below the cation ring with the change in the ethyl conformation, however, significant changes take place close to the H2 atom. With the ethyl group out of the plane, there is a reduction of the steric hindrance for anions approaching in the same plane of the ring (the blue region at right side is smaller in

Figure 7.d than 7.b), but at the same time there is an increase of free energy (less favorable) in front of the C2-H2 bond in comparison to the planar conformation. The calculation for the system without the restraints (Figure 7.e and f) presents again an intermediate behavior between the two cases with restricted conformations.

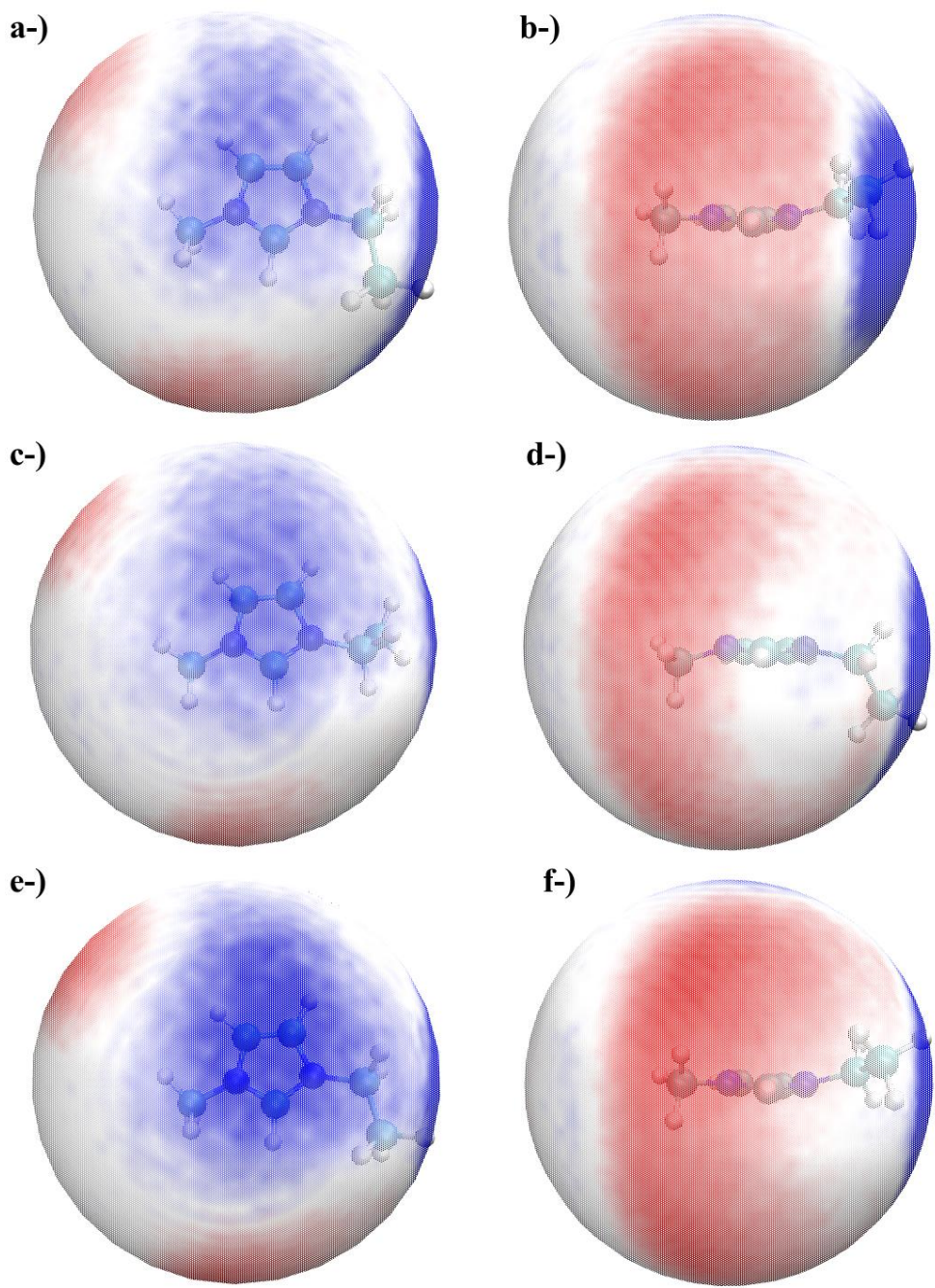

**Figure 7.** Free energy surfaces for [N(CN)$_2$]$^-$ around [C$_2$C$_1$im]$^+$ on a spherical shell of radius 5.55 Å around the N3 atom of the cation calculated by MD simulations of the liquid phase. Two different views are shown of the surface of occurrence of anions for each $[C_2C_1im]^+$ conformer: a-) and b-) simulation with $[C_2C_1im]^+$ restrained in planar conformation; c-) and d-) simulation with $[C_2C_1im]^+$ restrained at non-planar conformation; e-) and f-) simulation without dihedral restraint. The blue regions correspond to high values of free energy, and the red regions correspond to low values.

The results for the isolated dimer agree with the liquid simulations in the sense that the most favorable region changes from the front to the side of C2-H2 bond when the ethyl group is changed from planar to non-planar conformation. However, while in the liquid phase the most favorable region for the non-planar case was in the side of the methyl group, for the isolated dimer it is instead in the side of the ethyl group. This difference between the liquid and the isolated dimer behavior for the non-planar conformer is likely to be due to the dispersion interactions between the anion and the bulkier ethyl group in the isolated dimer that gives a gain in the interaction when compared to the situation close to the methyl group. In the liquid phase, on the other hand, there are others neighbor cations that also present dispersion interactions with the anion even if it is close to the methyl of the reference cation.

Summing up, there is a synergistic effect between flexibility of the ethyl group and the diffuse distribution of anions around the cation that difficult the crystallization of $[C_2C_1Im][N(CN)_2]$. As pointed out by Zhan *et al.*,[62] the occurrence of many local structures implies a multiplicity of local minimum in a landscape of energies. The shallow and broad megabasins in the energy landscape determine the ionic mobility, low viscosity, and low melting points of ionic liquids. The fact that anion–cation arrangement is further influenced by the flexibility of ethyl chain of $[C_2C_1Im]^+$ may affect both the melting point and the kinetics of crystallization. In other words, the ionic interactions and structural motifs that lead to large increase of entropy during fusion of the crystal, by the same token burden the crystallization and imply the glass forming ability of $[C_2C_1Im][N(CN)_2]$.

## V. CONCLUSIONS

The ionic liquid [$C_2C_1$im][N(CN)$_2$] is indeed easily supercooled and experiences glass transition as found in previous calorimetric studies.[31-35] However, we showed that in the same temperature of cold-crystallization observed after heating the glass ($T_{cc}$ ~ 210 K), crystallization during cooling can be observed if the sample is kept in isothermal condition for enough time. The same crystalline phase is obtained either by cold-crystallization or during cooling. The Raman spectra clearly indicate that [$C_2C_1$im][N(CN)$_2$] exhibits polymorphism: the low- and the high-temperature crystals have [$C_2C_1$im]$^+$ in planar and non-planar conformations, respectively. This solid-solid transition was not observed in some previous calorimetric studies most probably because the high-temperature crystal is formed closed to the melting temperature. The reason for the glass-forming ability of [$C_2C_1$im][N(CN)$_2$], even though being a low-viscous liquid and the ions being relatively simple, was pursued in the local structures calculated for ionic pairs by DFT or for the liquid phase by MD simulation. It is shown that the occurrence of anions around cations resulting from the MD simulation mirrors the free energy surface obtained from the quantum chemistry calculations of ionic pairs. We proposed that the delocalized occurrence of anions around the cation, together with the conformational flexibility of the ethyl group of the cation, hampers the location of ions in the definite positions needed for crystal growth. In other words, the structural motif that favors large entropy of the liquid phase, by the same token makes difficult crystallization along cooling of [$C_2C_1$im][N(CN)$_2$]. The broadly delocalized positions of [N(CN)$_2$]$^-$ around the cation, that is coupled to the conformational flexibility of the ethyl group of [$C_2C_1$im]$^+$, leads to the glass-forming ability of the ionic liquid [$C_2C_1$im][N(CN)$_2$].

## SUPPORTING INFORMATION

Infrared spectra of [$C_2C_1$im][N(CN)$_2$], the spherical grid used in the quantum chemistry calculation of the free energy surfaces, Raman spectra of mixed crystals of [$C_2C_1$im][N(CN)$_2$], Raman spectra in the region of C≡N stretching modes, radial distribution function and distribution of dihedral angles calculated by MD simulation.

## ACKNOWLEDGMENTS

The authors are indebted to FAPESP (Grants 2019/04785-9, 2017/12063-8, 2016/21070-5, 2014/15049-8) and CNPq (Grant 301553/2017-3) for financial support. We thank Prof. Flávio C. G. Gandra (Physics Institute, UNICAMP) for the calorimetric and X-ray diffraction measurements. We also thank the "Laboratório Nacional de Computação Científica – LNCC/MCTI, Brazil" for the use of the supercomputer SDumont (https://sdumont.lncc.br).

**SUPPORTING INFORMATION**

# Low Temperature Phase Transitions of the Ionic Liquid
# 1-Ethyl-3-methylimidazolium Dicyanamide

Kalil Bernardino, Thamires A. Lima, Mauro C. C. Ribeiro

*Laboratório de Espectroscopia Molecular, Departamento de Química Fundamental,*

*Instituto de Química, Universidade de São Paulo, 05508-000 São Paulo, SP, Brazil*

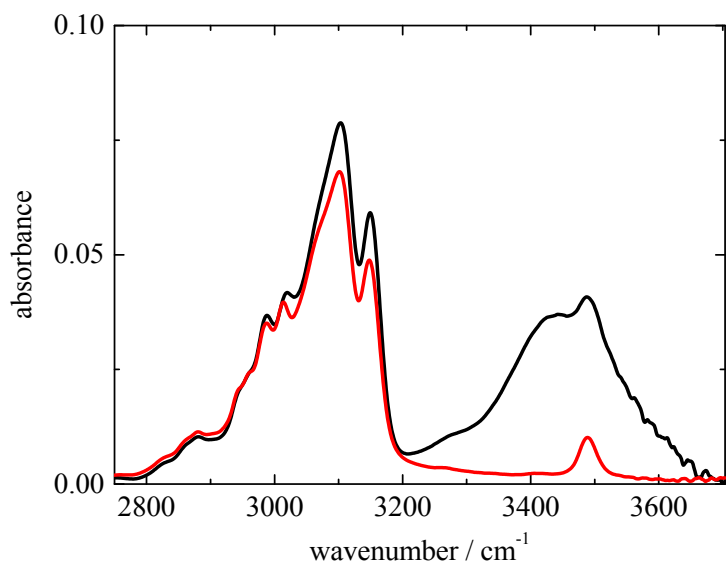

**Figure S1.** Infrared spectra of [C$_2$C$_1$im][N(CN)$_2$] obtained in ATR mode (attenuated total reflection) in a FT-IR spectrometer Vertex 80v, Bruker. The black line is the IR spectrum of a sample that was exposed to moisture for *ca.* 1 h. The red line is the IR spectrum for the same sample after leaving it for *ca.* 1 hour in the sample chamber of the spectrometer, which is kept under vacuum (~1 mbar). The absence of the broad band covering the 3400–3600 cm$^{-1}$ region indicates that water has been removed from the ionic liquid under vacuum. The spectral feature at ~3500 cm$^{-1}$ is a combination band of the [N(CN)$_2$]$^-$ anion.

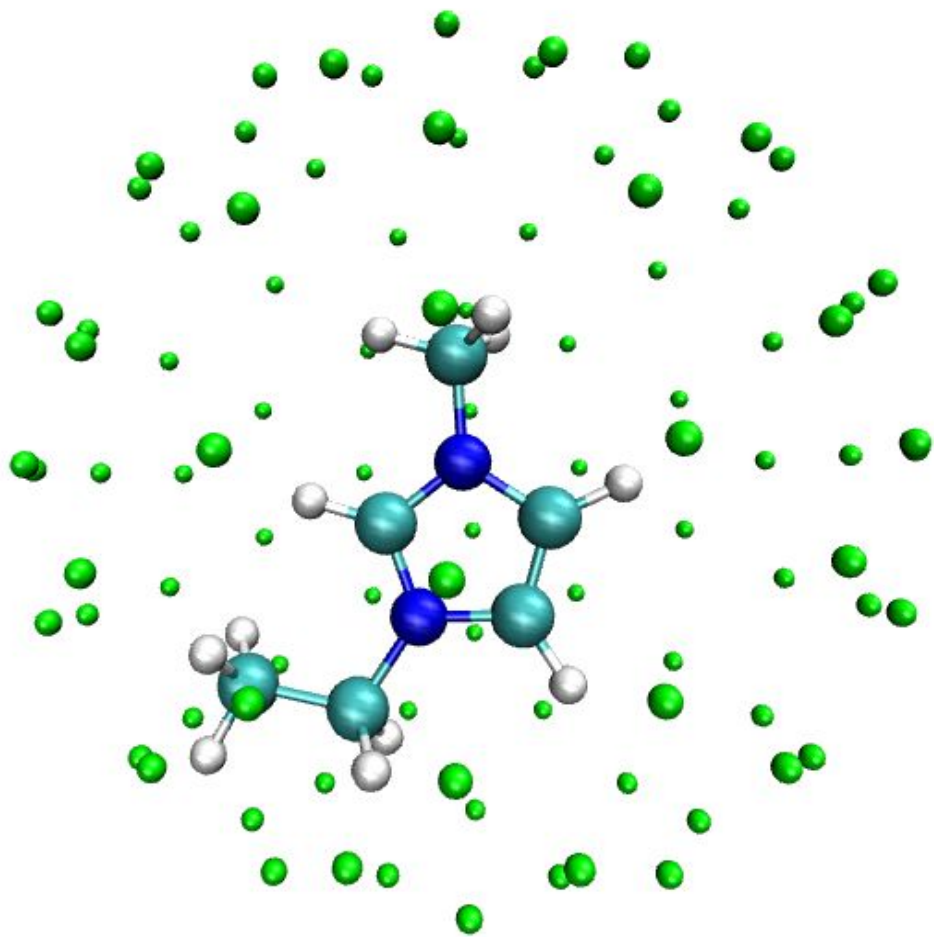

**Figure S2.** The 92 points spherical grid employed in the calculation of the free energy surface for the ionic pair. Every point is located at the same distance from the nitrogen N3 of the cation (for atom numbering, see Figure 1 in the manuscript). The anion [N(CN)$_2$]$^-$ is set up with its central nitrogen atom at one of these green points, and precessions and rotations are performed in order to produce an ensemble of structures for each translational point.

a)

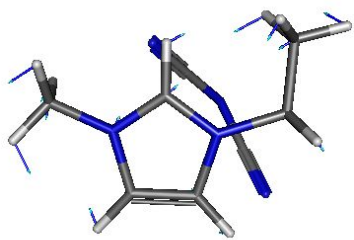

379 cm$^{-1}$

b)

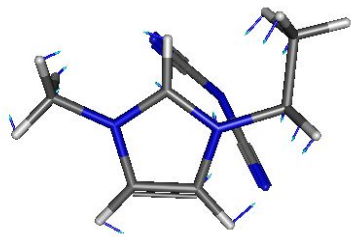

462 cm$^{-1}$

c)

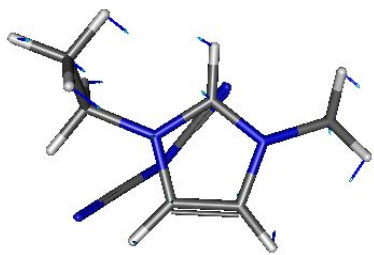

403 cm$^{-1}$

d)

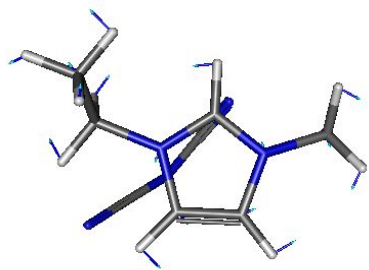

439 cm$^{-1}$

**Figure S3.** Atomic displacements obtained by DFT CAMB3LYP/Def2-TZVPD calculations of ion pairs for the vibrational modes related to the Raman bands shown in Figure 4. Figures a) and b) belong to the planar [C$_2$C$_1$im]$^+$ conformer; figures c) and d) belong to the non-planar [C$_2$C$_1$im]$^+$ conformer. There was no scaling factor applied to the calculated frequencies given in the figure.

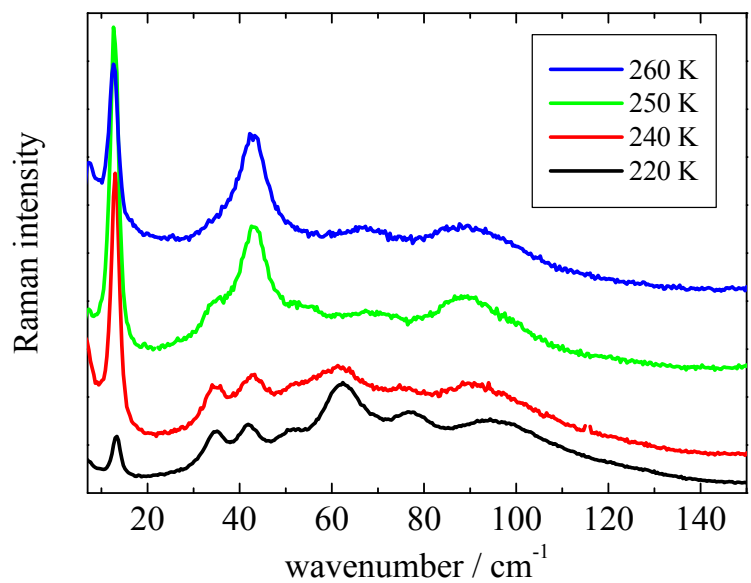

**Figure S4.** Low-frequency Raman spectra obtained along heating of [C$_2$C$_1$im][N(CN)$_2$] in the temperatures indicated in the figure. The Raman spectra correspond to mixtures of crystals I and II.

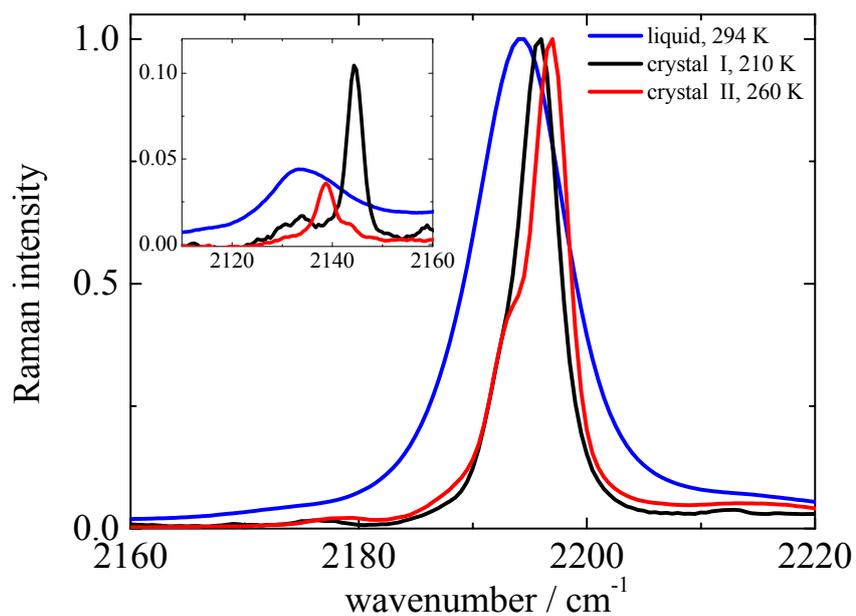

**Figure S5.** Raman spectra of $[C_2C_1im][N(CN)_2]$ in the different thermodynamic states indicated in the figure in the region of the symmetric C≡N stretching mode of the anion, $\nu_s(CN)$. The inset shows the region of the antisymmetric $\nu_{as}(CN)$ in a enlarged intensity scale. The spectra have been normalized by the maximum of intensity.

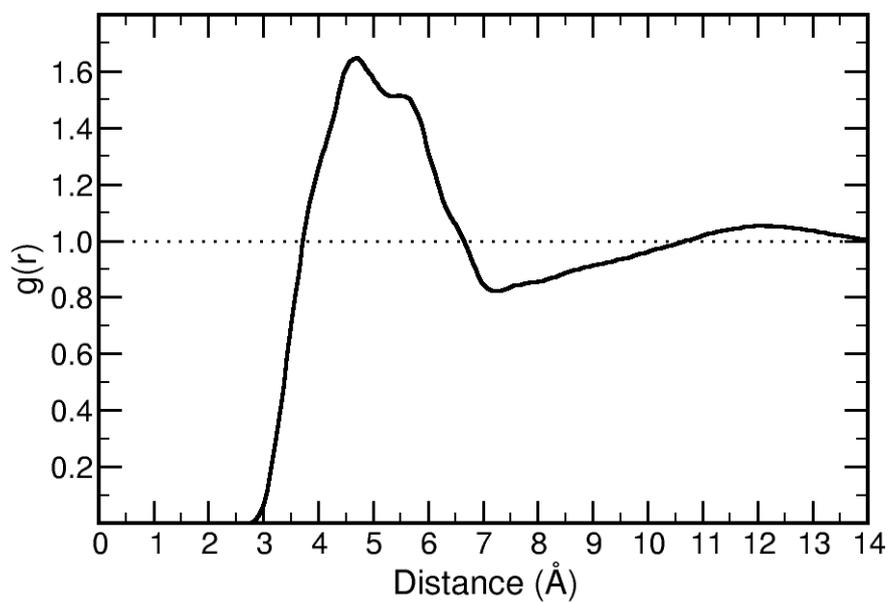

**Figure S6.** Radial distribution function of the central nitrogen of the anion around nitrogen N3 of the cation (for atom numbering, see Figure 1 in the manuscript).

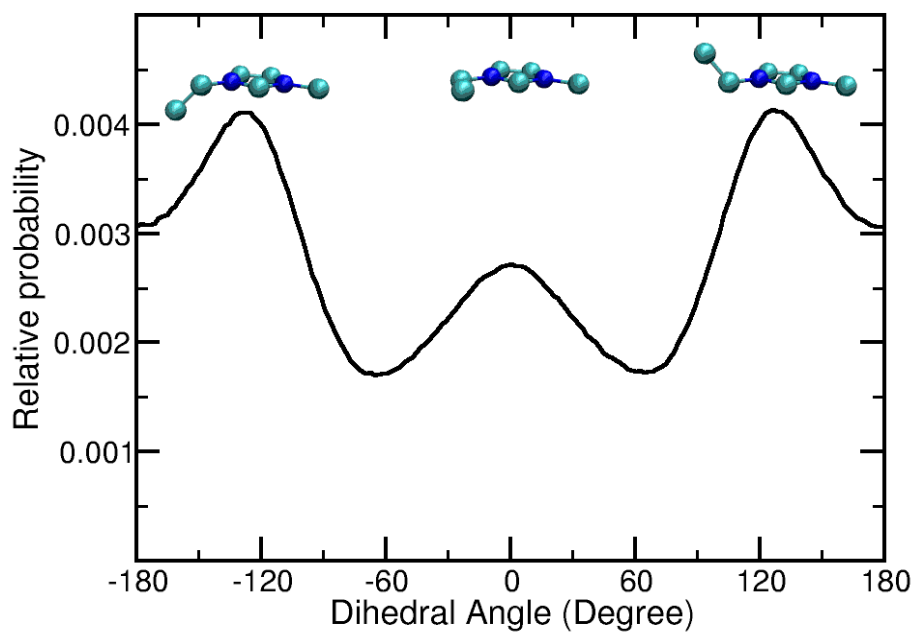

**Figure S7.** Distribution of the dihedral defined by the atoms C2 and N1 of the imidazolium ring and the two carbons of the ethyl group (for atom numbering, see Figure 1 in the manuscript). The structures of the conformations corresponding to maxima of probability are represented at the top without the hydrogen atoms for better visualization.